\documentclass[aps,prb,reprint,superscriptaddress,longbibliography]{revtex4-1}

\usepackage{graphicx}
\usepackage{epstopdf}
\usepackage{amsmath}
\usepackage{amsfonts}
\usepackage{amssymb}
\usepackage[a4paper=true,pagebackref=false]{hyperref}
\usepackage{color}
\usepackage{bm}

\begin{document}

\title{Defect-free SnTe topological crystalline insulator nanowires grown by molecular beam epitaxy on graphene}

\author{Janusz Sadowski}
\email[]{janusz.sadowski@ifpan.edu.pl}
\affiliation{Institute of Physics, Polish Academy of Sciences, Aleja Lotnikow 32/46, PL-02668 Warszawa, Poland}
\affiliation{Department of Physics and Electrical Engineering, Linnaeus University, SE-391 82 Kalmar, Sweden}
\affiliation{MAX-IV laboratory, Lund University, P.O. Box 118, SE-221 00 Lund, Sweden}

\author{Piotr Dziawa}
\author{Anna Kaleta}
\author{Bogus\l awa Kurowska}
\author{Anna Reszka}
\author{Tomasz Story}
\author{S\l awomir Kret}
\email[]{kret@ifpan.edu.pl}
\affiliation{Institute of Physics, Polish Academy of Sciences, Aleja Lotnikow 32/46, PL-02668 Warszawa, Poland} 

\begin{abstract}
SnTe topological crystalline insulator nanowires have been grown by molecular beam epitaxy on graphene/SiC substrates. The~nanowires have cubic rock-salt structure, they grow along [001] crystallographic direction and have four sidewalls consisting of \{100\} crystal planes known to host metallic surface states with Dirac dispersion. Thorough high resolution transmission electron microscopy investigations show that the nanowires grow on graphene in the van der Walls epitaxy mode induced when the catalyzing Au~nanoparticle mixes with Sn delivered from SnTe flux, providing liquid Au-Sn alloy. The~nanowires are totally free from structural defects, but their \{001\} sidewalls are prone to oxidation, which points out on necessity of depositing protective capping in view of exploiting the magneto-electric transport phenomena involving charge carriers occupying topologically protected states.\end{abstract}

\maketitle

\section{Introduction} 

SnTe belongs to a very well-known family of narrow bandgap IV-VI semiconductors, which have been studied because of their interesting electronic properties \citep{Wei1997} and applications, \textit{e.g.}~in thermoelectric energy harvesting and as mid-infrared detectors and emitters. \citep{Khokhlov2002} Quite recently SnTe and some of its ternary (multinary) alloys with Se and/or Pb were identified as topological crystalline insulators (TCI). \citep{Fu2011} Indeed IV-VI TCIs host topologically protected Dirac surface states in the vicinity of the distinct points of the Brillouin zone (in this case projection of the L points on the surface Brillouin zone). These states occur due to the coexistence of specific crystalline symmetries, \textit{i.e.}\ \{110\} mirror-plane of rock-salt crystal structure and electronic band inversion of a corresponding bulk crystal. \citep{Hsieh2012,Dziawa2012,Xu2012} The topological protection has been experimentally verified for two high symmetry surfaces of SnTe, namely \{100\} and \{111\}, first for \{100\} surfaces of bulk crystals, either after cleaving (in high vacuum) or for naturally grown crystal facets \citep{Tanaka2012,Tanaka2013,Dybko2017} and then for the respective surfaces of thin films grown by molecular beam epitaxy (MBE) \citep{Taskin2014, Zhang2014} or other methods such as hot-wall epitaxy and e-beam evaporation. \citep{Polley2016,Zeljkovic2015,Walkup2018} With epitaxially grown crystalline layers it is possible to access the topologically protected states on surfaces which are not easy cleavage planes, only reachable when bulk crystals are used. Moreover in thin layer geometry, at some thickness limit, the wavefunctions of topologically protected carriers (Dirac states) start to overlap \citep{Li2017} leading to new phenomena, \textit{e.g.}\ occurrence of two-dimensional topological crystalline insulator and quantum spin Hall insulator in the same material. \citep{Safaei2015} In this letter we report on an interesting kind of low-dimensional structures fabricated from SnTe TCI material, quasi one dimensional nanowires (NWs) deposited on graphene/SiC(6H). The~graphene-TI heterostructures are foreseen to exhibit interesting properties due to combination of high mobility 2D graphene charge carriers and the spin-texture of carriers in a TI material. \citep{Song2018} To our knowledge the MBE growth of SnTe NWs has not been reported before neither on graphene nor on any other substrate. Several literature reports on SnTe NWs and nanocrystals published already \citep{Li2013,Safdar2013,Safdar2014,Shen2014,Zou2015} concern the nanostructures obtained \textit{via} chemical vapor deposition technique which cannot compete with MBE in terms of controlling the NW geometrical parameters (lengths and diameters) and the ability to grow NW heterostructures. \citep{Mata2013}

\section{Experimental methods}

\newcommand*{\degr}{$^\circ$}

The NWs have been grown by MBE on graphene/SiC(6H) substrates using gold assisted  growth mode. Graphene was deposited on a C-face of SiC(0001) by Chemical Vapor Deposition technique, according to the procedure described in Ref.~\citenum{Strupinski2011}. As the source of gold nanocrystals the 10~nm Au nanoparticles (NP) suspended in {\bf H2O} (BBI solutions) have been used. After wetting graphene/SiC substrate with Au nanoparticles solution the sample was dried using gaseous nitrogen, mounted on the MBE sample holder and introduced to the load lock of the MBE system. After initial degassing at about 200~\degr C the Au(NP)/graphene/SiC substrate was heated to about 700~\degr C, then the substrate temperature was decreased to 400~\degr C -- 440~\degr C and the deposition of SnTe was started using compound SnTe source (Knudsen effusion cell) with beam equivalent pressure of about $4.5 \times 10^{-8}$~mbar. The~growth has been monitored by reflection high energy electron diffraction (RHEED) system. Nanoscale morphologies of the as-grown samples have been investigated by the scanning electron microscope (Hitachi SU-70) operating at acceleration voltage of 15~kV. Structural properties of NWs and of the SnTe/graphene/SiC interface have been investigated with FEI-Titan 80-300 transmission electron microscope operating at 300~kV, equipped with an image corrector. For TEM investigations NWs are transferred to a holey carbon film suported by a Cu mesh mechanicaly, without using any liquid with dispersed NWs. The~Helios Nanolab 600 FIB was used to prepare the transparent for electrons SnTe/graphene/SiC interface cross-section by means of focused ion beam techique and lift-off  procedure with Oniprobe nanomanipulator. The~electron deposited platinum was used first to protect NWs then the thicker ion deposited layer was supplied. The~low loss spectra were aquired in STEM mode using The~Quantum 966 Gatan Image Filter with energy resolution of about 0.95~eV. EDS spectroscopy was performed using EDAX system. All the TEM investigations have been done at room temperature.

\section{Results and discussion}
\begin{figure}[h]
\centering
	\includegraphics[width=0.99\columnwidth]{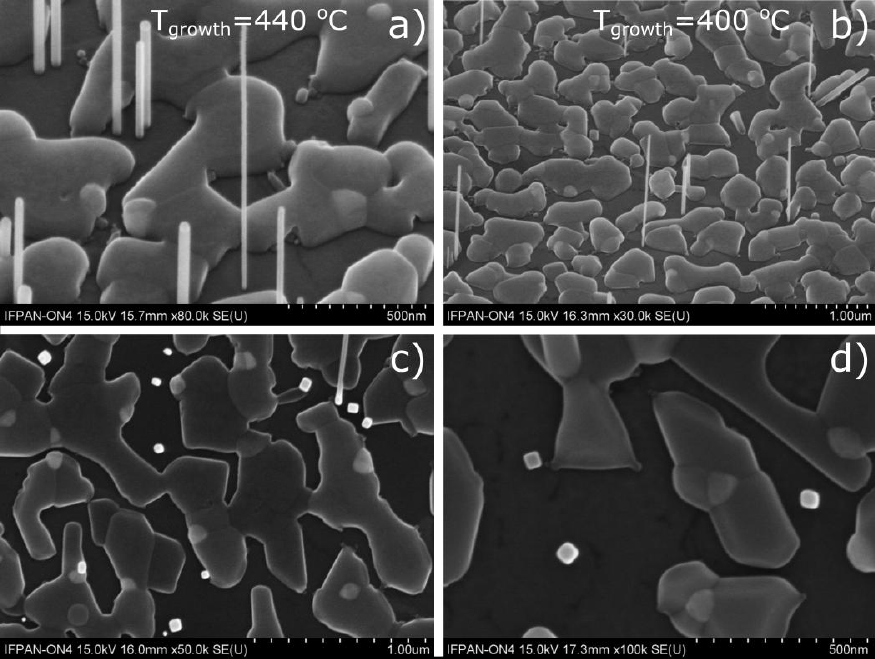}
	\caption{SEM images of Au-catalyzed SnTe nanowires deposited on graphene/SiC(6H) at 440~\degr C (a), (c); and at 400~\degr C (b), (d). Upper panels -- sample plane tilted at 45\degr\ angle with respect to the e-beam direction; lower panels -- sample plane perpendicular to the e-beam direction. Top of the NWs are visible in this projection as bright square-like objects.}
	\label{fgr:sem}
\end{figure}

Figure~\ref{fgr:sem} shows scanning electron microscopy (SEM) images of SnTe deposited on graphene/SiC substrate at the substrate temperature of 440~\degr C (sample 1) and 400~\degr C (sample 2).

\begin{figure}[h]
\centering
	\includegraphics[width=0.99\columnwidth]{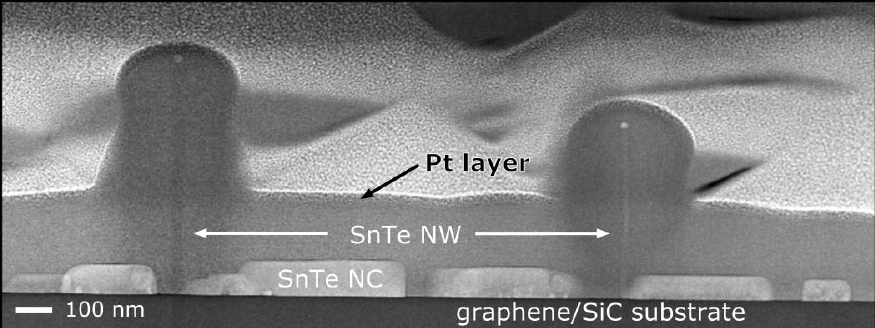}
	\caption{Low resolution TEM image of the Focused Ion Beam cross-section of sample (1). The~NWs are embedded in electron deposited platinum layer.}
	\label{fgr:fib}
\end{figure}

The majority of NWs visible in Fig.~\ref{fgr:sem} are about 0.5~$\mu$m long and 50~nm thick, occasionally much longer and thinner NWs occur. Apparently, beside the SnTe NWs, the graphene/SiC substrate is also partially covered with planar SnTe. The~Au nanocrystals which coagulate and do not catalyze SnTe NWs are visible as brighter spots at the irregular planar SnTe deposits. In spite of slightly lower surface NW densities there are no significant differences in the structure of both samples, but the planar MBE growth of SnTe is slightly more pronounced for sample (2) grown at lower substrate temperature than sample~(1).   

SEM images taken with e-beam perpendicular to the substrate (parallel to the NWs) -- Figs.~\ref{fgr:sem}c and~\ref{fgr:sem}d indicate that NWs have square cross sections, which corresponds to [001] NW growth axes. We have observed previously that, in contrast to III-V and II-VI semiconductor NWs which usually grow along [111] crystallographic direction, the IV-VI NWs naturally grow along [001] direction, \citep{Dziawa2010} which implies their 4-fold axial symmetry and square cross section.

In Figs.~\ref{fgr:sem}c and~\ref{fgr:sem}d it can be seen that there is lack of correlation of azimuthal orientation of the NWs. However almost all the NWs grow perpendicularly to the substrate. Figure~\ref{fgr:fib} shows focused ion beam (FIB) cross-section acquired from sample (1). Two SnTe NWs embedded in thick platinum layer can be identified in the image. 

In the FIB cross-section also the blocks of planar SnTe nanocrystals (NC) are visible. All the  NCs have flat surfaces parallel to the interface plane (shown in more details in Fig.~\ref{fgr:fib}). The~two SnTe NWs indicated on the image are perpendicular to the substrate plane.

\begin{figure*}[t!]
\centering
	\includegraphics[width=0.99\textwidth]{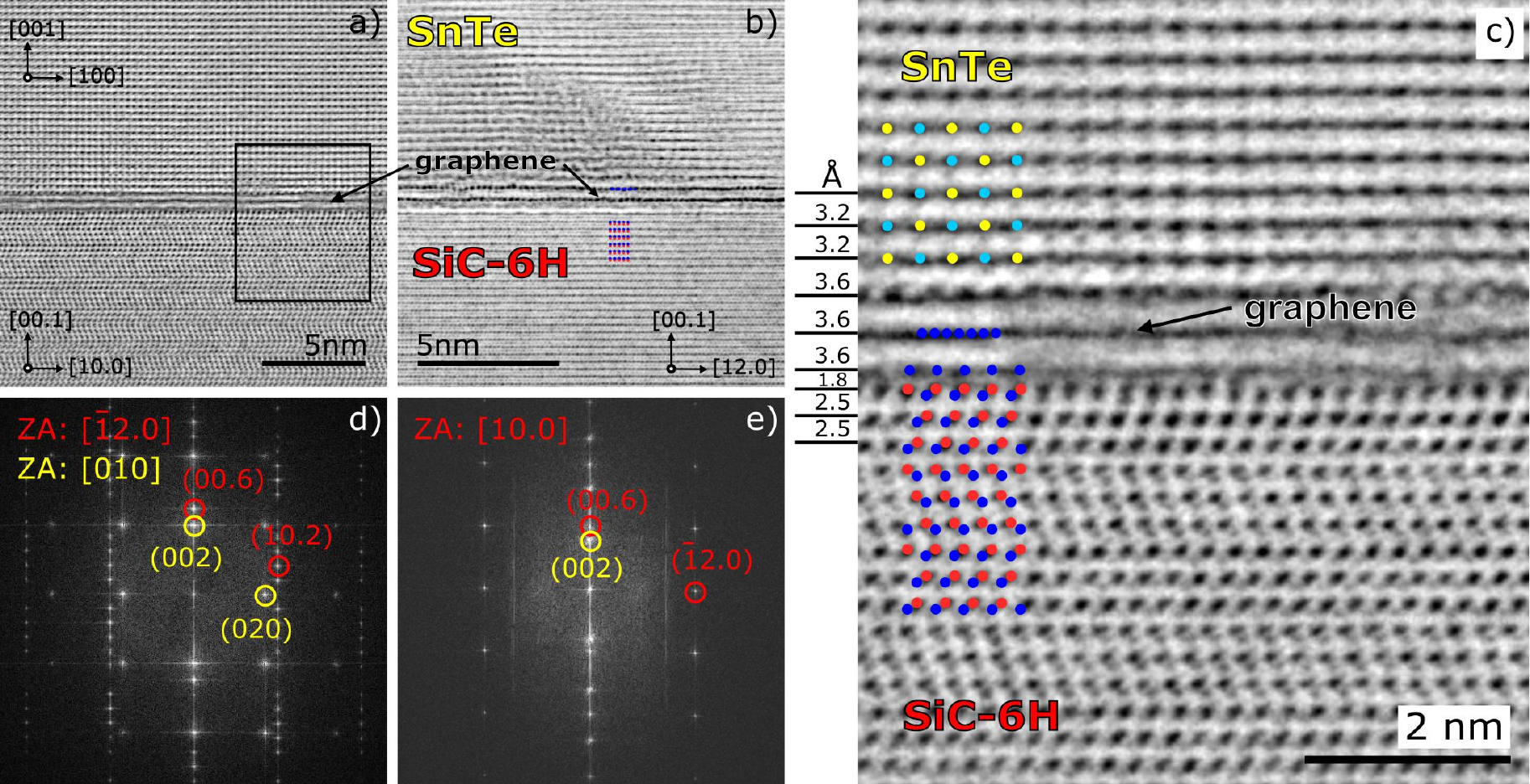}
	\caption{(a) -- (c) Inversed contrast (atoms are black), NCSI high resolution TEM images of the SnTe/graphene/SiC(6H) interface. The~positions of Si and C atoms in graphene/SiC are indicated by red and blue dots, respectively; whereas those of Sn, and Te in SnTe -- by light-blue and yellow dots; (d) and (e) -- corresponding Fast Fourier Transform (FFT) images of the same region as shown in (c), with indexed FFT features origination from SiC(6H) (red) and SnTe (yellow).}
	\label{fgr:interface}
\end{figure*}

Figure~\ref{fgr:interface} shows high resolution HRTEM image of the FIB cross section of SnTe/graphene/SiC hetero-interface (the same specimen measured in different zone axes). The~negative spherical aberration imaging (NCSI) mode \citep{Jia2010} helps in visualizing the carbon atoms in SiC as well as in the graphene interlayer. With the spherical aberration correction settings used ($\rm{Cs}=-40$~$\mu$m) the atomic columns in SiC and graphene appear bright against darker background. Due to the low density of NWs we were unable to find a cross section with SnTe(NW)/substrate interface in the specimen suitable for HRTEM investigations. Only planar SnTe NCs could be identified. 

In the interface region with graphene bi-layer we observe the distances between the lattice planes which differ from both SiC and SnTe (see Fig.~\ref{fgr:interface}c). Analyzing thoroughly the HRTEM image of this interface we can conclude that SiC is terminated by the carbon layer spaced 1.8~\AA\ from the Si-plane of SiC(6H), similarly to that reported by~\citeauthor{Norimatsu2009}. \citep{Norimatsu2009} Next we observe 3.6~\AA\ distance between two graphene planes and the first monolayer of SnTe which is directly placed over the second layer of graphene. The~SnTe crystal shown in Fig.~\ref{fgr:interface} is slightly twisted around [001] direction so the atomic structure of SnTe is not perfectly resolved. The~visible degradation of graphene interlayer under electron beam was detected after 5~minute exposition to 300~kV electrons and total amorphization appeared after about 20~minutes in the normal HRTEM observation conditions at high magnification. 

Even though the cross section shown in Fig.~\ref{fgr:interface} applies to the interface between planar SnTe and graphene/SiC(6H) substrate we believe that the interface between SnTe NW and graphene/SiC(6H) looks the same since similarly to planar SnTe NCs the rock-salt SnTe NWs grow along [001] direction. The~planar growth of very thin SnTe (1~to~4~ML) on graphene/SiC substrate has been investigated previously in the context of the robust in-plane SnTe ferroelectric properties. \citep{Chang2016} The authors of \mbox{Ref.~\citenum{Chang2016}} also observed only (001) orientation of thin SnTe islands deposited by MBE on graphene/SiC. On the other hand for SnTe nanostructures (nanoplates and planar nanowires) deposited on Silicon by CVD technique both (100) and (111) planar SnTe growth have been observed.\citep{Shen2014}

In spite of the dissimilarities between (001) planes of cubic SnTe, with 4-fold symmetry and the 6-fold symmetry of the graphene/SiC(6H) substrate there is some degree of correlation between SnTe(100) and graphene/SiC(6H), evidenced in the FFT features (see Fig.~\ref{fgr:interface}d). In the case of self-catalyzed GaAs NWs grown of graphene \citep{Munshi2012} such correlation was even more pronounced and concerned also azimuthal correlations between GaAs NWs with [111] growth axes and graphene, which is not observed in our case. For GaAs[111] NWs both the NWs and the substrate share common 6-fold symmetries, which does not occur for (001)SnTe and graphene/SiC(6H).  

\begin{figure}[h]
\centering
	\includegraphics[width=0.99\columnwidth]{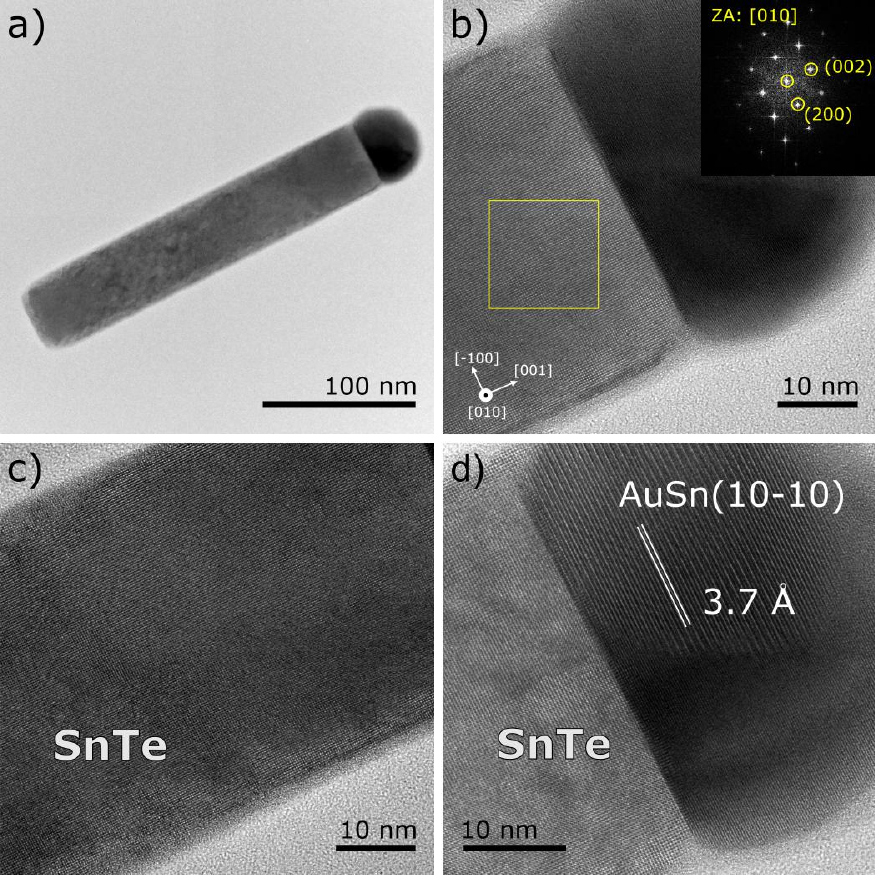}
	\caption{(a) low resolution TEM image of a nanowire taken from sample (2); (b) high resolution image of the top nanowire part. The~inset at the upper right corner shows 2D Fast Fourier Transform patterns taken from the region indicated by the yellow frame. \mbox{The~4-fold} symmetry FFT patterns correspond to the cubic rock-salt structure of the NW; (c) top part of the SnTe NW below the catalyst nanocrystal, the scale bar corresponds to 10~nm; (d) the solidified droplet at the NW tip.}
	\label{fgr:tem}
\end{figure}

Figure~\ref{fgr:tem} shows TEM images of a nanowire picked up from sample (2). It can be seen that the surface of the upper NW part (under the catalyst droplet) is smooth as opposed to the bottom where we observe pronounced surface roughness due to oxidation.

\begin{figure}[h]
\centering
	\includegraphics[width=0.99\columnwidth]{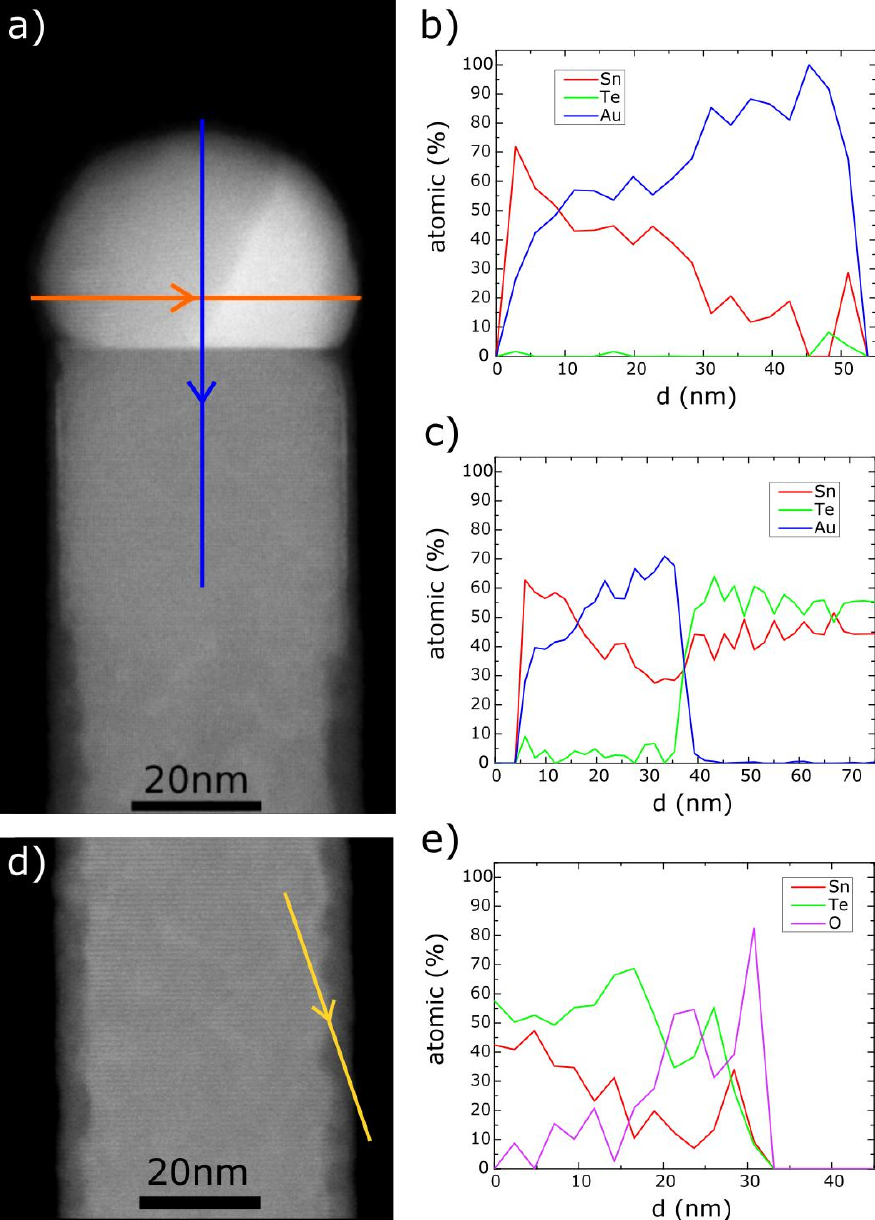}
	\caption{(a) STEM image of solidified droplet at the SnTe nanowire tip. EDS elemental composition scans across (b) perpendicular and (c) parallel to the NW axis taken along blue and orange arrows, respectively; (d) STEM image of the bottom part of the same NW with the yellow arrow showing EDS line scan across the sidewall of NW; (e) the elemental profiles of Sn, Te and O.}
	\label{fgr:eds}
\end{figure}

The catalyst nanocrystal at the NW tip shown in Fig.~\ref{fgr:tem}d is inhomogeneous; the part visible at the upper section of the image is Sn-rich, as evidenced by the EDS composition scans described further in the text. The~spacing between the crystallographic planes clearly visible in the upper part of the solidified droplet corresponds well to 3.74~\AA\ distance between the (10-10) planes of the hexagonal Au$_{0.5}$Sn$_{0.5}$ phase. \citep{Arora2014}   

As indicated above the SnTe NWs with [001] growth axes are totally free from structural defects. In the NWs shown in Fig.~\ref{fgr:tem} and in many other NWs picked up from samples (1) and (2) the apparent inhomogeneity of the crystalline structure of a solidified catalyst droplet can be seen. The~energy dispersion X-ray spectroscopy (EDS) elemental composition scans performed across the droplet (along the orange and blue lines shown in Figs.~\ref{fgr:eds}a show that the larger part of the droplet is Sn-rich. 
The~part of the solidified droplet represented with darker contrast in the TEM image is Sn rich, with the Sn content changing from 100\% to 20\% along the perpendicular scan line and from 60\% to 30\% along the parallel scan line. The~scan directions are indicated by the arrows in~Fig.~\ref{fgr:eds}a.
In this part the atomic planes perpendicular to the NW axes could be identified (see Fig.~\ref{fgr:tem}). 

Thorough inspection of TEM images of the top part of SnTe NW shown in Fig.~\ref{fgr:eds} reveals that the NW sidewalls are oxidized (see the bottom part of Fig.~\ref{fgr:eds}a and Fig.~\ref{fgr:eds}d). Amorphous oxide is manifested by slightly darker contrast of the close-to-sidewall region, with respect to that of the NW body. It has been also confirmed by the EDS scan through this region (Fig.~\ref{fgr:eds}e). Apparently only the very top part of the NW -- down to about 40~nm below the interface with the catalyst droplet is free from oxidation. In some NWs this oxidation is very pronounced, but in all the cases the very top NW parts are protected against oxidation by a thin (2~--~3~ML) film of Au, or Au-Sn alloy (this region is too thin to unequivocally determine its composition). The~problem of SnTe oxidation has recently been investigated by~\citeauthor{Berchenko2018} \citep{Berchenko2018} The~authors of Ref.~\citenum{Berchenko2018} show that the amorphous oxide at the SnTe(100) surfaces consists of the mixture of Sn and Te oxides and builds up during the first 10~minutes exposure of SnTe to air, then it's thickness increases slowly in the time-scale of months, and can reach up to 20~nm after 1~--~2~years.  In the NWs studied by us we have noticed different degrees of oxidation of NWs picked up from different samples, which have been stored in air for similar time. We cannot directly infer what causes the differences in the oxidation rate; tentatively we attribute that to different stoichiometries of SnTe NWs. An example of NW's oxidation presented in Fig.~\ref{fgr:eds}d and the corresponding EDS profile (Fig.~\ref{fgr:eds}e) shows significant contribution of oxygen at the sidewall. The~sidewalls of NWs collected from sample (1) grown at 440~\degr C are much more oxidized than those collected from sample (2), grown at 400~\degr C; and due to the known thermodynamic properties of SnTe \citep{Brebrick1964} we can infer that sample (1) is more Sn-rich than sample (2).

\begin{figure}[h!]
\centering
	\includegraphics[width=1.00\columnwidth]{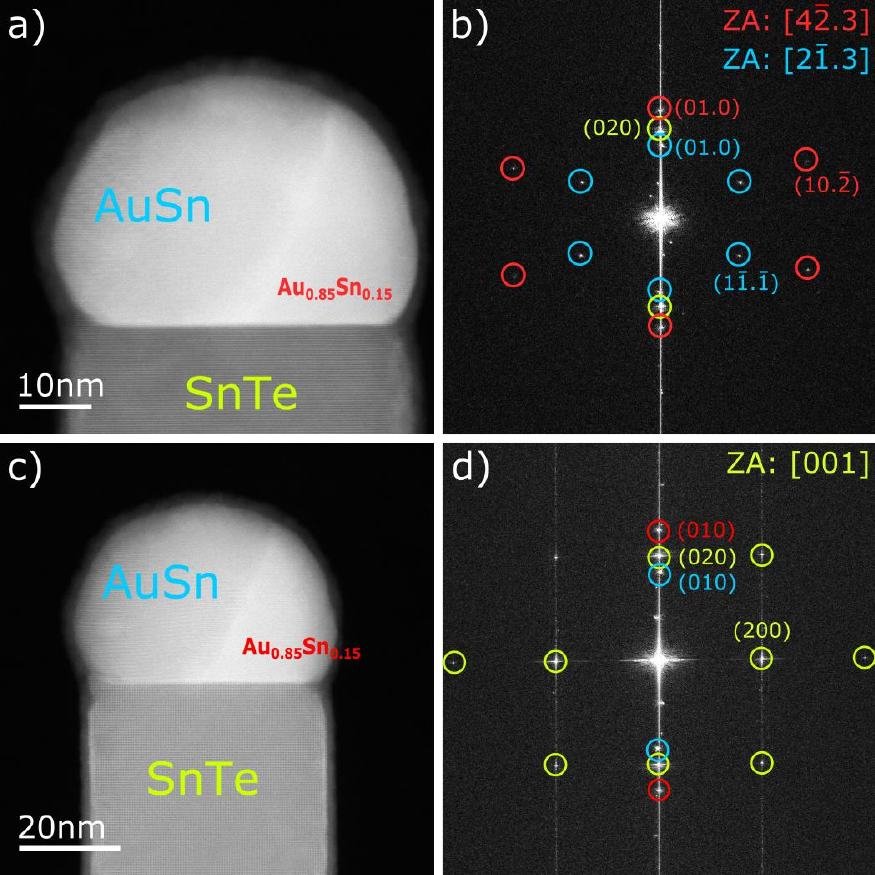}
	\caption{STEM-HAADF images -- (a), (b) and corresponding 2D FFT images (c), (d) of different areas of the top part of SnTe nanowire in different zone axes: (c) -- [4-2.3] AuSn and [2-1.3] Au$_{0.85}$Sn$_{0.15}$, (d) [001] SnTe. The~2D FFT patterns corresponding to SnTe, Au$_{0.5}$Sn$_{0.5}$ and Au$_{0.85}$Sn$_{0.15}$ are marked with yellow, light blue and red colors, respectively.}
	\label{fgr:stem}
\end{figure}

\begin{figure}[h!]
\centering
	\includegraphics[width=1.00\columnwidth]{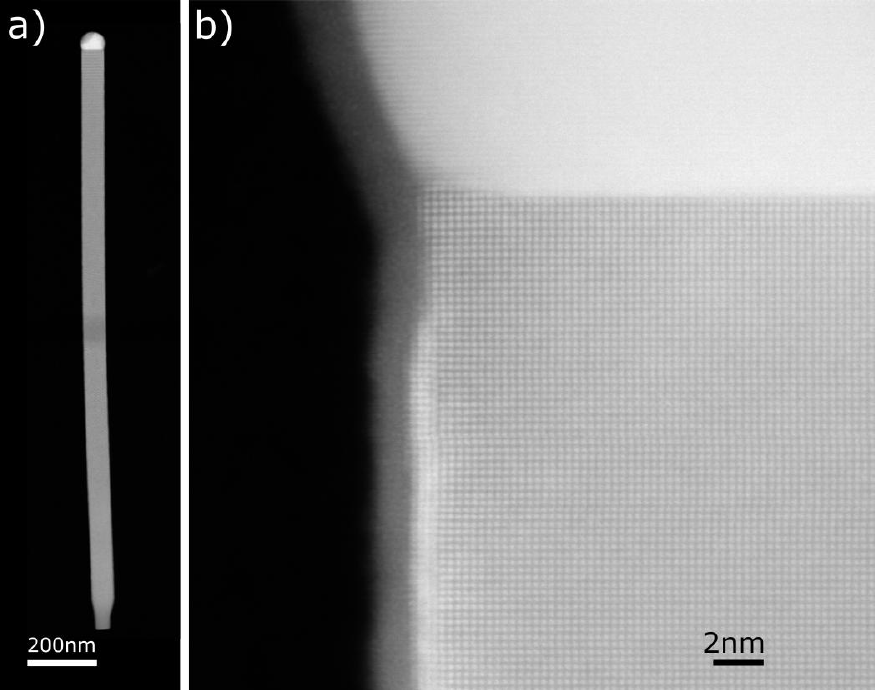}
	\caption{STEM-HAADF  images of the same SnTe NW at different magnifications.}
	\label{fgr:long}
\end{figure}

As can be seen in Figs.~\ref{fgr:eds}b and~\ref{fgr:eds}c the Au rich part of the droplet at the NW tip still contains substantial amount of Sn. To identify which phase of the Au-Sn alloy is present in the droplet we have analyzed the two-dimensional Fast Fourier Transformation (2D FFT) images of the different NW tip and droplet areas (see Fig.~\ref{fgr:stem}). The~2D FFT patterns obtained in three different zone axis directions correspond well to: Au$_{0.5}$Sn$_{0.5}$ -- the main part of the droplet (light blue) and Au$_{0.85}$Sn$_{0.15}$ (red). The~patterns corresponding to SnTe in the NW body are marked with yellow. Both Au-Sn intermetallic compounds are liquid at the SnTe NWs growth temperature applied here (400~\degr C -- 440~\degr C), \citep{Okamoto1984} which indicates that the metal nanoparticle at the NW tip is in the liquid state during the NWs growth. 

Figure~\ref{fgr:long} shows two STEM-HAADF images of the SnTe NW grown on graphene/SiC collected from the sample (3) grown at the substrate temperature of 450 \degr C for much longer time (3h) than samples (1) and (2),~(1~h). Here the NWs are longer; however they have similar diameters as those picked up from samples (1) and (2) (see Figs.~\ref{fgr:sem},~\ref{fgr:fib} and~\ref{fgr:tem}) and are also totally free from structural defects across the whole NW lengths. 

\begin{figure}[t!]
\centering
	\includegraphics[width=1.00\columnwidth]{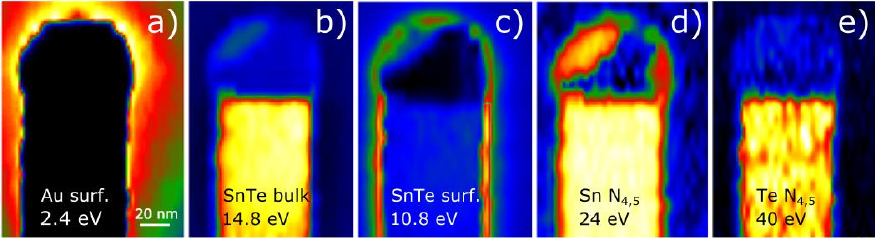}
	\caption{(a)--(c) -- Color coded maps of plasmon excitations of the top part of SnTe nanowire detected by EELS. (a) -- Au surface plasmons (2.4~eV); (b) -- SnTe bulk plasmons (14.8~eV); (c) SnTe surface plasmons (10.8~eV);  (d)--(e) -- color coded maps of the low loss edges of: (d)~Sn~N$_{4,5}$ edge (24~eV), (e)~Te~N$_{4,5}$ edge (40~eV) jump ratios visualizing distribution of Sn and Te in the top part of the SnTe nanowire. The spurious fluctuations of Sn and Te signals in panels (d) and (e) are experimental artifacts due to the low signal to noise ratios.}
	\label{fgr:eels}
\end{figure}

Similarly to the NW shown in Figs.~\ref{fgr:tem}--\ref{fgr:stem}, here the NW axis is also parallel to SnTe[001] crystallographic direction. Likewise, the solidified metal droplet at the NW tip has inhomogeneous composition. 

As can clearly be seen in Fig.~\ref{fgr:long}b the outer parts of the NW sidewall edges appear slightly brighter than the NW body (visible in all the SnTe NWs inspected by us). EDS elemental composition scans of this region prove that similarly to the NW shown in Figs.~\ref{fgr:tem}--\ref{fgr:stem}, in these regions the NW sidewalls are rich in gold and consist of a~very thin (2~--~3~ML) film of gold or gold-tin intermetallic alloy. As~mentioned before, this very thin Au-rich shell prevents the SnTe NW sidewall from oxidation, which appears as light NW sidewalls necking visible in the lower part of the image shown in Fig.~\ref{fgr:sem}a. 

Complementary information about the composition of the top part of SnTe NWs has been obtained by detection of plasmon resonances with use of the electron energy loss spectroscopy (EELS) in STEM. EELS is a powerful technique enabling detections of electron energy losses due to excitation in a broad spectral range from optical energies up to vacuum ultraviolet. \citep{Colliex2016} Figure~\ref{fgr:eels} shows the 2D maps of EELS intensity obtained by integration of signal for different energy windows spectra of the SnTe NW tip and solidified Au-Sn droplet registered for the~NW shown in Fig.~\ref{fgr:long} (lower energy part of the EELS spectrum), and Sn and Te element distribution mapping due to the signal corresponding to N$_{4,5}$ edges of both elements. The~plasmon-loss spectra are detected for energies corresponding to Au localized surface plasmons (2.4~eV), \citep{Bosman2007} SnTe volume plasmons (14.8~eV) and SnTe surface plasmons (10.8~eV). \citep{Cook1971} The width of the EELS signal integration window was set at 1~eV, 3.4~eV and 3~eV, respectively. 

In accordance with EDS and structural information obtained previously, the plasmon resonances observed in EELS confirm the presence of the thin Au-rich film at the top parts of the NW sidewalls. The~higher energy part of the EELS spectrum (in the 20~eV -- 50~eV range) enables direct elemental mapping of the NW and the metal catalyst at the top. The~2D mapping of the EELS signal corresponding to absorption at Sn~N$_{4,5}$ (24 eV) and Te~N$_{4,5}$~(40~eV) edges (see Figs.~\ref{fgr:eels}d and~\ref{fgr:eels}e) confirms the inhomogeneous composition of the catalyzing Au-Sn nanoparticle (solid at the TEM measurements temperature), and the negligible amount of Te in it.

\section{Conclusions}

In summary -- we have investigated the Au-assisted MBE growth of SnTe topological crystalline insulator nanowires on graphene/SiC(6H) substrate. The~NWs have cubic rock-salt structure and grow along [001] axes which implies their square cross-sections and occurrence of four \{100\} sidewalls recognized as topologically protected surfaces of the SnTe topological crystalline insulator. Analysis of the metal catalyst nanocrystals at the NW tips reveals that they contain substantial amount of Sn. Two distinct phases of Au-Sn intermetallic alloy: Au$_{0.5}$Sn$_{0.5}$ and Au$_{0.85}$Sn$_{0.15}$ have been identified in the solidified catalyst droplet. Both phases are liquid at the NW growth temperature (400~\degr C -- 440~\degr C) which indicates that the NWs grow in the VLS mode. In the majority of NWs, thoroughly investigated by TEM, no structural defects have been found along the entire NW lengths. In most of the NWs we have observed the oxidation of the NW sidewalls. Only the very top parts of the NWs 40~nm -- 80~nm below the catalyzing metal nanoparticle are not oxidized due to protection by the very thin Au-rich film. This points out on necessity of depositing protective shells on the SnTe NW sidewalls in view of studying the transport phenomena involving charge carriers occupying topologically protected states residing at four SnTe(100) NW sidewalls.

\section*{Acknowledgements}
This work has been supported by the research projects No.~2014/13/B/ST3/04489, and 2014/15/B/ST3/03833 financed by the National Science Centre (Poland). The~authors thank the Institute of Electronic Materials Technology and Dr.~W.~Strupi\'nski for graphene/SiC substrates.

\bibliography{Defect-free-SnTe-TCI-NWs}

\begin{thebibliography}{36}%
\makeatletter
\providecommand \@ifxundefined [1]{%
 \@ifx{#1\undefined}
}%
\providecommand \@ifnum [1]{%
 \ifnum #1\expandafter \@firstoftwo
 \else \expandafter \@secondoftwo
 \fi
}%
\providecommand \@ifx [1]{%
 \ifx #1\expandafter \@firstoftwo
 \else \expandafter \@secondoftwo
 \fi
}%
\providecommand \natexlab [1]{#1}%
\providecommand \enquote  [1]{``#1''}%
\providecommand \bibnamefont  [1]{#1}%
\providecommand \bibfnamefont [1]{#1}%
\providecommand \citenamefont [1]{#1}%
\providecommand \href@noop [0]{\@secondoftwo}%
\providecommand \href [0]{\begingroup \@sanitize@url \@href}%
\providecommand \@href[1]{\@@startlink{#1}\@@href}%
\providecommand \@@href[1]{\endgroup#1\@@endlink}%
\providecommand \@sanitize@url [0]{\catcode `\\12\catcode `\$12\catcode
  `\&12\catcode `\#12\catcode `\^12\catcode `\_12\catcode `\%12\relax}%
\providecommand \@@startlink[1]{}%
\providecommand \@@endlink[0]{}%
\providecommand \url  [0]{\begingroup\@sanitize@url \@url }%
\providecommand \@url [1]{\endgroup\@href {#1}{\urlprefix }}%
\providecommand \urlprefix  [0]{URL }%
\providecommand \Eprint [0]{\href }%
\providecommand \doibase [0]{http://dx.doi.org/}%
\providecommand \selectlanguage [0]{\@gobble}%
\providecommand \bibinfo  [0]{\@secondoftwo}%
\providecommand \bibfield  [0]{\@secondoftwo}%
\providecommand \translation [1]{[#1]}%
\providecommand \BibitemOpen [0]{}%
\providecommand \bibitemStop [0]{}%
\providecommand \bibitemNoStop [0]{.\EOS\space}%
\providecommand \EOS [0]{\spacefactor3000\relax}%
\providecommand \BibitemShut  [1]{\csname bibitem#1\endcsname}%
\let\auto@bib@innerbib\@empty
\bibitem [{\citenamefont {Wei}\ and\ \citenamefont {Zunger}(1997)}]{Wei1997}%
  \BibitemOpen
  \bibfield  {author} {\bibinfo {author} {\bibfnamefont {Su-Huai}\ \bibnamefont
  {Wei}}\ and\ \bibinfo {author} {\bibfnamefont {Alex}\ \bibnamefont
  {Zunger}},\ }\bibfield  {title} {\enquote {\bibinfo {title} {Electronic and
  structural anomalies in lead chalcogenides},}\ }\href {\doibase
  10.1103/PhysRevB.55.13605} {\bibfield  {journal} {\bibinfo  {journal} {Phys.
  Rev. B}\ }\textbf {\bibinfo {volume} {55}},\ \bibinfo {pages} {13605--13610}
  (\bibinfo {year} {1997})}\BibitemShut {NoStop}%
\bibitem [{\citenamefont {Khokhlov}(2002)}]{Khokhlov2002}%
  \BibitemOpen
  \bibfield  {author} {\bibinfo {author} {\bibfnamefont {D.}~\bibnamefont
  {Khokhlov}},\ }\href {https://books.google.pl/books?id=y4P4Kf399l8C} {\emph
  {\bibinfo {title} {Lead Chalcogenides: Physics and Applications}}},\
  Optoelectronic properties of semiconductors and superlattices\ (\bibinfo
  {publisher} {Taylor \& Francis},\ \bibinfo {year} {2002})\BibitemShut
  {NoStop}%
\bibitem [{\citenamefont {Fu}(2011)}]{Fu2011}%
  \BibitemOpen
  \bibfield  {author} {\bibinfo {author} {\bibfnamefont {Liang}\ \bibnamefont
  {Fu}},\ }\bibfield  {title} {\enquote {\bibinfo {title} {Topological
  crystalline insulators},}\ }\href {\doibase 10.1103/PhysRevLett.106.106802}
  {\bibfield  {journal} {\bibinfo  {journal} {Phys. Rev. Lett.}\ }\textbf
  {\bibinfo {volume} {106}},\ \bibinfo {pages} {106802} (\bibinfo {year}
  {2011})}\BibitemShut {NoStop}%
\bibitem [{\citenamefont {Hsieh}\ \emph {et~al.}(2012)\citenamefont {Hsieh},
  \citenamefont {Lin}, \citenamefont {Liu}, \citenamefont {Duan}, \citenamefont
  {Bansil},\ and\ \citenamefont {Fu}}]{Hsieh2012}%
  \BibitemOpen
  \bibfield  {author} {\bibinfo {author} {\bibfnamefont {Timothy~H.}\
  \bibnamefont {Hsieh}}, \bibinfo {author} {\bibfnamefont {Hsin}\ \bibnamefont
  {Lin}}, \bibinfo {author} {\bibfnamefont {Junwei}\ \bibnamefont {Liu}},
  \bibinfo {author} {\bibfnamefont {Wenhui}\ \bibnamefont {Duan}}, \bibinfo
  {author} {\bibfnamefont {Arun}\ \bibnamefont {Bansil}}, \ and\ \bibinfo
  {author} {\bibfnamefont {Liang}\ \bibnamefont {Fu}},\ }\bibfield  {title}
  {\enquote {\bibinfo {title} {Topological crystalline insulators in the snte
  material class},}\ }\href {http://dx.doi.org/10.1038/ncomms1969} {\bibfield
  {journal} {\bibinfo  {journal} {Nature Communications}\ }\textbf {\bibinfo
  {volume} {3}},\ \bibinfo {pages} {982} (\bibinfo {year} {2012})}\BibitemShut
  {NoStop}%
\bibitem [{\citenamefont {Dziawa}\ \emph {et~al.}(2012)\citenamefont {Dziawa},
  \citenamefont {Kowalski}, \citenamefont {Dybko}, \citenamefont {Buczko},
  \citenamefont {Szczerbakow}, \citenamefont {Szot}, \citenamefont
  {Lusakowska}, \citenamefont {Balasubramanian}, \citenamefont {Wojek},
  \citenamefont {Berntsen}, \citenamefont {Tjernberg},\ and\ \citenamefont
  {Story}}]{Dziawa2012}%
  \BibitemOpen
  \bibfield  {author} {\bibinfo {author} {\bibfnamefont {P.}~\bibnamefont
  {Dziawa}}, \bibinfo {author} {\bibfnamefont {B.~J.}\ \bibnamefont
  {Kowalski}}, \bibinfo {author} {\bibfnamefont {K.}~\bibnamefont {Dybko}},
  \bibinfo {author} {\bibfnamefont {R.}~\bibnamefont {Buczko}}, \bibinfo
  {author} {\bibfnamefont {A.}~\bibnamefont {Szczerbakow}}, \bibinfo {author}
  {\bibfnamefont {M.}~\bibnamefont {Szot}}, \bibinfo {author} {\bibfnamefont
  {E.}~\bibnamefont {Lusakowska}}, \bibinfo {author} {\bibfnamefont
  {T.}~\bibnamefont {Balasubramanian}}, \bibinfo {author} {\bibfnamefont
  {B.~M.}\ \bibnamefont {Wojek}}, \bibinfo {author} {\bibfnamefont {M.~H.}\
  \bibnamefont {Berntsen}}, \bibinfo {author} {\bibfnamefont {O.}~\bibnamefont
  {Tjernberg}}, \ and\ \bibinfo {author} {\bibfnamefont {T.}~\bibnamefont
  {Story}},\ }\bibfield  {title} {\enquote {\bibinfo {title} {Topological
  crystalline insulator states in pb1-xsnxse},}\ }\href
  {http://dx.doi.org/10.1038/nmat3449} {\bibfield  {journal} {\bibinfo
  {journal} {Nature Materials}\ }\textbf {\bibinfo {volume} {11}},\ \bibinfo
  {pages} {1023} (\bibinfo {year} {2012})}\BibitemShut {NoStop}%
\bibitem [{\citenamefont {Xu}\ \emph {et~al.}(2012)\citenamefont {Xu},
  \citenamefont {Liu}, \citenamefont {Alidoust}, \citenamefont {Neupane},
  \citenamefont {Qian}, \citenamefont {Belopolski}, \citenamefont {Denlinger},
  \citenamefont {Wang}, \citenamefont {Lin}, \citenamefont {Wray},
  \citenamefont {Landolt}, \citenamefont {Slomski}, \citenamefont {Dil},
  \citenamefont {Marcinkova}, \citenamefont {Morosan}, \citenamefont {Gibson},
  \citenamefont {Sankar}, \citenamefont {Chou}, \citenamefont {Cava},
  \citenamefont {Bansil},\ and\ \citenamefont {Hasan}}]{Xu2012}%
  \BibitemOpen
  \bibfield  {author} {\bibinfo {author} {\bibfnamefont {Su-Yang}\ \bibnamefont
  {Xu}}, \bibinfo {author} {\bibfnamefont {Chang}\ \bibnamefont {Liu}},
  \bibinfo {author} {\bibfnamefont {N.}~\bibnamefont {Alidoust}}, \bibinfo
  {author} {\bibfnamefont {M.}~\bibnamefont {Neupane}}, \bibinfo {author}
  {\bibfnamefont {D.}~\bibnamefont {Qian}}, \bibinfo {author} {\bibfnamefont
  {I.}~\bibnamefont {Belopolski}}, \bibinfo {author} {\bibfnamefont {J.~D.}\
  \bibnamefont {Denlinger}}, \bibinfo {author} {\bibfnamefont {Y.~J.}\
  \bibnamefont {Wang}}, \bibinfo {author} {\bibfnamefont {H.}~\bibnamefont
  {Lin}}, \bibinfo {author} {\bibfnamefont {L.~A.}\ \bibnamefont {Wray}},
  \bibinfo {author} {\bibfnamefont {G.}~\bibnamefont {Landolt}}, \bibinfo
  {author} {\bibfnamefont {B.}~\bibnamefont {Slomski}}, \bibinfo {author}
  {\bibfnamefont {J.~H.}\ \bibnamefont {Dil}}, \bibinfo {author} {\bibfnamefont
  {A.}~\bibnamefont {Marcinkova}}, \bibinfo {author} {\bibfnamefont
  {E.}~\bibnamefont {Morosan}}, \bibinfo {author} {\bibfnamefont
  {Q.}~\bibnamefont {Gibson}}, \bibinfo {author} {\bibfnamefont
  {R.}~\bibnamefont {Sankar}}, \bibinfo {author} {\bibfnamefont {F.~C.}\
  \bibnamefont {Chou}}, \bibinfo {author} {\bibfnamefont {R.~J.}\ \bibnamefont
  {Cava}}, \bibinfo {author} {\bibfnamefont {A.}~\bibnamefont {Bansil}}, \ and\
  \bibinfo {author} {\bibfnamefont {M.~Z.}\ \bibnamefont {Hasan}},\ }\bibfield
  {title} {\enquote {\bibinfo {title} {Observation of a topological crystalline
  insulator phase and topological phase transition in pb1-xsnxte},}\ }\href
  {http://dx.doi.org/10.1038/ncomms2191} {\bibfield  {journal} {\bibinfo
  {journal} {Nature Communications}\ }\textbf {\bibinfo {volume} {3}},\
  \bibinfo {pages} {1192} (\bibinfo {year} {2012})}\BibitemShut {NoStop}%
\bibitem [{\citenamefont {Tanaka}\ \emph {et~al.}(2012)\citenamefont {Tanaka},
  \citenamefont {Ren}, \citenamefont {Sato}, \citenamefont {Nakayama},
  \citenamefont {Souma}, \citenamefont {Takahashi}, \citenamefont {Segawa},\
  and\ \citenamefont {Ando}}]{Tanaka2012}%
  \BibitemOpen
  \bibfield  {author} {\bibinfo {author} {\bibfnamefont {Y.}~\bibnamefont
  {Tanaka}}, \bibinfo {author} {\bibfnamefont {Zhi}\ \bibnamefont {Ren}},
  \bibinfo {author} {\bibfnamefont {T.}~\bibnamefont {Sato}}, \bibinfo {author}
  {\bibfnamefont {K.}~\bibnamefont {Nakayama}}, \bibinfo {author}
  {\bibfnamefont {S.}~\bibnamefont {Souma}}, \bibinfo {author} {\bibfnamefont
  {T.}~\bibnamefont {Takahashi}}, \bibinfo {author} {\bibfnamefont {Kouji}\
  \bibnamefont {Segawa}}, \ and\ \bibinfo {author} {\bibfnamefont {Yoichi}\
  \bibnamefont {Ando}},\ }\bibfield  {title} {\enquote {\bibinfo {title}
  {Experimental realization of a topological crystalline insulator in snte},}\
  }\href {http://dx.doi.org/10.1038/nphys2442} {\bibfield  {journal} {\bibinfo
  {journal} {Nature Physics}\ }\textbf {\bibinfo {volume} {8}},\ \bibinfo
  {pages} {800} (\bibinfo {year} {2012})}\BibitemShut {NoStop}%
\bibitem [{\citenamefont {Tanaka}\ \emph {et~al.}(2013)\citenamefont {Tanaka},
  \citenamefont {Shoman}, \citenamefont {Nakayama}, \citenamefont {Souma},
  \citenamefont {Sato}, \citenamefont {Takahashi}, \citenamefont {Novak},
  \citenamefont {Segawa},\ and\ \citenamefont {Ando}}]{Tanaka2013}%
  \BibitemOpen
  \bibfield  {author} {\bibinfo {author} {\bibfnamefont {Y.}~\bibnamefont
  {Tanaka}}, \bibinfo {author} {\bibfnamefont {T.}~\bibnamefont {Shoman}},
  \bibinfo {author} {\bibfnamefont {K.}~\bibnamefont {Nakayama}}, \bibinfo
  {author} {\bibfnamefont {S.}~\bibnamefont {Souma}}, \bibinfo {author}
  {\bibfnamefont {T.}~\bibnamefont {Sato}}, \bibinfo {author} {\bibfnamefont
  {T.}~\bibnamefont {Takahashi}}, \bibinfo {author} {\bibfnamefont
  {M.}~\bibnamefont {Novak}}, \bibinfo {author} {\bibfnamefont {Kouji}\
  \bibnamefont {Segawa}}, \ and\ \bibinfo {author} {\bibfnamefont {Yoichi}\
  \bibnamefont {Ando}},\ }\bibfield  {title} {\enquote {\bibinfo {title} {Two
  types of dirac-cone surface states on the (111) surface of the topological
  crystalline insulator snte},}\ }\href {\doibase 10.1103/PhysRevB.88.235126}
  {\bibfield  {journal} {\bibinfo  {journal} {Phys. Rev. B}\ }\textbf {\bibinfo
  {volume} {88}},\ \bibinfo {pages} {235126} (\bibinfo {year}
  {2013})}\BibitemShut {NoStop}%
\bibitem [{\citenamefont {Dybko}\ \emph {et~al.}(2017)\citenamefont {Dybko},
  \citenamefont {Szot}, \citenamefont {Szczerbakow}, \citenamefont {Gutowska},
  \citenamefont {Zajarniuk}, \citenamefont {Domagala}, \citenamefont
  {Szewczyk}, \citenamefont {Story},\ and\ \citenamefont
  {Zawadzki}}]{Dybko2017}%
  \BibitemOpen
  \bibfield  {author} {\bibinfo {author} {\bibfnamefont {K.}~\bibnamefont
  {Dybko}}, \bibinfo {author} {\bibfnamefont {M.}~\bibnamefont {Szot}},
  \bibinfo {author} {\bibfnamefont {A.}~\bibnamefont {Szczerbakow}}, \bibinfo
  {author} {\bibfnamefont {M.~U.}\ \bibnamefont {Gutowska}}, \bibinfo {author}
  {\bibfnamefont {T.}~\bibnamefont {Zajarniuk}}, \bibinfo {author}
  {\bibfnamefont {J.~Z.}\ \bibnamefont {Domagala}}, \bibinfo {author}
  {\bibfnamefont {A.}~\bibnamefont {Szewczyk}}, \bibinfo {author}
  {\bibfnamefont {T.}~\bibnamefont {Story}}, \ and\ \bibinfo {author}
  {\bibfnamefont {W.}~\bibnamefont {Zawadzki}},\ }\bibfield  {title} {\enquote
  {\bibinfo {title} {Experimental evidence for topological surface states
  wrapping around a bulk snte crystal},}\ }\href {\doibase
  10.1103/PhysRevB.96.205129} {\bibfield  {journal} {\bibinfo  {journal} {Phys.
  Rev. B}\ }\textbf {\bibinfo {volume} {96}},\ \bibinfo {pages} {205129}
  (\bibinfo {year} {2017})}\BibitemShut {NoStop}%
\bibitem [{\citenamefont {Taskin}\ \emph {et~al.}(2014)\citenamefont {Taskin},
  \citenamefont {Yang}, \citenamefont {Sasaki}, \citenamefont {Segawa},\ and\
  \citenamefont {Ando}}]{Taskin2014}%
  \BibitemOpen
  \bibfield  {author} {\bibinfo {author} {\bibfnamefont {A.~A.}\ \bibnamefont
  {Taskin}}, \bibinfo {author} {\bibfnamefont {Fan}\ \bibnamefont {Yang}},
  \bibinfo {author} {\bibfnamefont {Satoshi}\ \bibnamefont {Sasaki}}, \bibinfo
  {author} {\bibfnamefont {Kouji}\ \bibnamefont {Segawa}}, \ and\ \bibinfo
  {author} {\bibfnamefont {Yoichi}\ \bibnamefont {Ando}},\ }\bibfield  {title}
  {\enquote {\bibinfo {title} {Topological surface transport in epitaxial snte
  thin films grown on bi${}_{2}$te${}_{3}$},}\ }\href {\doibase
  10.1103/PhysRevB.89.121302} {\bibfield  {journal} {\bibinfo  {journal} {Phys.
  Rev. B}\ }\textbf {\bibinfo {volume} {89}},\ \bibinfo {pages} {121302}
  (\bibinfo {year} {2014})}\BibitemShut {NoStop}%
\bibitem [{\citenamefont {Zhang}\ \emph {et~al.}(2014)\citenamefont {Zhang},
  \citenamefont {Baek}, \citenamefont {Ha}, \citenamefont {Zhang},
  \citenamefont {Wyrick}, \citenamefont {Davydov}, \citenamefont {Kuk},\ and\
  \citenamefont {Stroscio}}]{Zhang2014}%
  \BibitemOpen
  \bibfield  {author} {\bibinfo {author} {\bibfnamefont {Duming}\ \bibnamefont
  {Zhang}}, \bibinfo {author} {\bibfnamefont {Hongwoo}\ \bibnamefont {Baek}},
  \bibinfo {author} {\bibfnamefont {Jeonghoon}\ \bibnamefont {Ha}}, \bibinfo
  {author} {\bibfnamefont {Tong}\ \bibnamefont {Zhang}}, \bibinfo {author}
  {\bibfnamefont {Jonathan}\ \bibnamefont {Wyrick}}, \bibinfo {author}
  {\bibfnamefont {Albert~V.}\ \bibnamefont {Davydov}}, \bibinfo {author}
  {\bibfnamefont {Young}\ \bibnamefont {Kuk}}, \ and\ \bibinfo {author}
  {\bibfnamefont {Joseph~A.}\ \bibnamefont {Stroscio}},\ }\bibfield  {title}
  {\enquote {\bibinfo {title} {Quasiparticle scattering from topological
  crystalline insulator snte (001) surface states},}\ }\href {\doibase
  10.1103/PhysRevB.89.245445} {\bibfield  {journal} {\bibinfo  {journal} {Phys.
  Rev. B}\ }\textbf {\bibinfo {volume} {89}},\ \bibinfo {pages} {245445}
  (\bibinfo {year} {2014})}\BibitemShut {NoStop}%
\bibitem [{\citenamefont {Polley}\ \emph {et~al.}(2016)\citenamefont {Polley},
  \citenamefont {Jovic}, \citenamefont {Su}, \citenamefont {Saghir},
  \citenamefont {Newby}, \citenamefont {Kowalski}, \citenamefont {Jakiela},
  \citenamefont {Barcz}, \citenamefont {Guziewicz}, \citenamefont
  {Balasubramanian}, \citenamefont {Balakrishnan}, \citenamefont {Laverock},\
  and\ \citenamefont {Smith}}]{Polley2016}%
  \BibitemOpen
  \bibfield  {author} {\bibinfo {author} {\bibfnamefont {C.~M.}\ \bibnamefont
  {Polley}}, \bibinfo {author} {\bibfnamefont {V.}~\bibnamefont {Jovic}},
  \bibinfo {author} {\bibfnamefont {T.-Y.}\ \bibnamefont {Su}}, \bibinfo
  {author} {\bibfnamefont {M.}~\bibnamefont {Saghir}}, \bibinfo {author}
  {\bibfnamefont {D.}~\bibnamefont {Newby}}, \bibinfo {author} {\bibfnamefont
  {B.~J.}\ \bibnamefont {Kowalski}}, \bibinfo {author} {\bibfnamefont
  {R.}~\bibnamefont {Jakiela}}, \bibinfo {author} {\bibfnamefont
  {A.}~\bibnamefont {Barcz}}, \bibinfo {author} {\bibfnamefont
  {M.}~\bibnamefont {Guziewicz}}, \bibinfo {author} {\bibfnamefont
  {T.}~\bibnamefont {Balasubramanian}}, \bibinfo {author} {\bibfnamefont
  {G.}~\bibnamefont {Balakrishnan}}, \bibinfo {author} {\bibfnamefont
  {J.}~\bibnamefont {Laverock}}, \ and\ \bibinfo {author} {\bibfnamefont
  {K.~E.}\ \bibnamefont {Smith}},\ }\bibfield  {title} {\enquote {\bibinfo
  {title} {Observation of surface states on heavily indium-doped snte(111), a
  superconducting topological crystalline insulator},}\ }\href {\doibase
  10.1103/PhysRevB.93.075132} {\bibfield  {journal} {\bibinfo  {journal} {Phys.
  Rev. B}\ }\textbf {\bibinfo {volume} {93}},\ \bibinfo {pages} {075132}
  (\bibinfo {year} {2016})}\BibitemShut {NoStop}%
\bibitem [{\citenamefont {Zeljkovic}\ \emph {et~al.}(2015)\citenamefont
  {Zeljkovic}, \citenamefont {Walkup}, \citenamefont {Assaf}, \citenamefont
  {Scipioni}, \citenamefont {Sankar}, \citenamefont {Chou},\ and\ \citenamefont
  {Madhavan}}]{Zeljkovic2015}%
  \BibitemOpen
  \bibfield  {author} {\bibinfo {author} {\bibfnamefont {Ilija}\ \bibnamefont
  {Zeljkovic}}, \bibinfo {author} {\bibfnamefont {Daniel}\ \bibnamefont
  {Walkup}}, \bibinfo {author} {\bibfnamefont {Badih~A.}\ \bibnamefont
  {Assaf}}, \bibinfo {author} {\bibfnamefont {Kane~L.}\ \bibnamefont
  {Scipioni}}, \bibinfo {author} {\bibfnamefont {R.}~\bibnamefont {Sankar}},
  \bibinfo {author} {\bibfnamefont {Fangcheng}\ \bibnamefont {Chou}}, \ and\
  \bibinfo {author} {\bibfnamefont {Vidya}\ \bibnamefont {Madhavan}},\
  }\bibfield  {title} {\enquote {\bibinfo {title} {Strain engineering dirac
  surface states in heteroepitaxial topological crystalline insulator thin
  films},}\ }\href {http://dx.doi.org/10.1038/nnano.2015.177} {\bibfield
  {journal} {\bibinfo  {journal} {Nature Nanotechnology}\ }\textbf {\bibinfo
  {volume} {10}},\ \bibinfo {pages} {849} (\bibinfo {year} {2015})}\BibitemShut
  {NoStop}%
\bibitem [{\citenamefont {Walkup}\ \emph {et~al.}(2018)\citenamefont {Walkup},
  \citenamefont {Assaf}, \citenamefont {Scipioni}, \citenamefont {Sankar},
  \citenamefont {Chou}, \citenamefont {Chang}, \citenamefont {Lin},
  \citenamefont {Zeljkovic},\ and\ \citenamefont {Madhavan}}]{Walkup2018}%
  \BibitemOpen
  \bibfield  {author} {\bibinfo {author} {\bibfnamefont {Daniel}\ \bibnamefont
  {Walkup}}, \bibinfo {author} {\bibfnamefont {Badih~A.}\ \bibnamefont
  {Assaf}}, \bibinfo {author} {\bibfnamefont {Kane~L.}\ \bibnamefont
  {Scipioni}}, \bibinfo {author} {\bibfnamefont {R.}~\bibnamefont {Sankar}},
  \bibinfo {author} {\bibfnamefont {Fangcheng}\ \bibnamefont {Chou}}, \bibinfo
  {author} {\bibfnamefont {Guoqing}\ \bibnamefont {Chang}}, \bibinfo {author}
  {\bibfnamefont {Hsin}\ \bibnamefont {Lin}}, \bibinfo {author} {\bibfnamefont
  {Ilija}\ \bibnamefont {Zeljkovic}}, \ and\ \bibinfo {author} {\bibfnamefont
  {Vidya}\ \bibnamefont {Madhavan}},\ }\bibfield  {title} {\enquote {\bibinfo
  {title} {Interplay of orbital effects and nanoscale strain in topological
  crystalline insulators},}\ }\href {\doibase 10.1038/s41467-018-03887-5}
  {\bibfield  {journal} {\bibinfo  {journal} {Nature Communications}\ }\textbf
  {\bibinfo {volume} {9}},\ \bibinfo {pages} {1550} (\bibinfo {year}
  {2018})}\BibitemShut {NoStop}%
\bibitem [{\citenamefont {Li}\ and\ \citenamefont {Niu}(2017)}]{Li2017}%
  \BibitemOpen
  \bibfield  {author} {\bibinfo {author} {\bibfnamefont {Xiao}\ \bibnamefont
  {Li}}\ and\ \bibinfo {author} {\bibfnamefont {Qian}\ \bibnamefont {Niu}},\
  }\bibfield  {title} {\enquote {\bibinfo {title} {Topological phase
  transitions in thin films by tuning multivalley boundary-state couplings},}\
  }\href {\doibase 10.1103/PhysRevB.95.241411} {\bibfield  {journal} {\bibinfo
  {journal} {Phys. Rev. B}\ }\textbf {\bibinfo {volume} {95}},\ \bibinfo
  {pages} {241411} (\bibinfo {year} {2017})}\BibitemShut {NoStop}%
\bibitem [{\citenamefont {Safaei}\ \emph {et~al.}(2015)\citenamefont {Safaei},
  \citenamefont {Galicka}, \citenamefont {Kacman},\ and\ \citenamefont
  {Buczko}}]{Safaei2015}%
  \BibitemOpen
  \bibfield  {author} {\bibinfo {author} {\bibfnamefont {S}~\bibnamefont
  {Safaei}}, \bibinfo {author} {\bibfnamefont {M}~\bibnamefont {Galicka}},
  \bibinfo {author} {\bibfnamefont {P}~\bibnamefont {Kacman}}, \ and\ \bibinfo
  {author} {\bibfnamefont {R}~\bibnamefont {Buczko}},\ }\bibfield  {title}
  {\enquote {\bibinfo {title} {Quantum spin hall effect in iv-vi topological
  crystalline insulators},}\ }\href
  {http://stacks.iop.org/1367-2630/17/i=6/a=063041} {\bibfield  {journal}
  {\bibinfo  {journal} {New Journal of Physics}\ }\textbf {\bibinfo {volume}
  {17}},\ \bibinfo {pages} {063041} (\bibinfo {year} {2015})}\BibitemShut
  {NoStop}%
\bibitem [{\citenamefont {Song}\ \emph {et~al.}(2018)\citenamefont {Song},
  \citenamefont {Soriano}, \citenamefont {Cummings}, \citenamefont {Robles},
  \citenamefont {Ordej{\'o}n},\ and\ \citenamefont {Roche}}]{Song2018}%
  \BibitemOpen
  \bibfield  {author} {\bibinfo {author} {\bibfnamefont {Kenan}\ \bibnamefont
  {Song}}, \bibinfo {author} {\bibfnamefont {David}\ \bibnamefont {Soriano}},
  \bibinfo {author} {\bibfnamefont {Aron~W.}\ \bibnamefont {Cummings}},
  \bibinfo {author} {\bibfnamefont {Roberto}\ \bibnamefont {Robles}}, \bibinfo
  {author} {\bibfnamefont {Pablo}\ \bibnamefont {Ordej{\'o}n}}, \ and\ \bibinfo
  {author} {\bibfnamefont {Stephan}\ \bibnamefont {Roche}},\ }\bibfield
  {title} {\enquote {\bibinfo {title} {Spin proximity effects in
  graphene/topological insulator heterostructures},}\ }\href {\doibase
  10.1021/acs.nanolett.7b05482} {\bibfield  {journal} {\bibinfo  {journal}
  {Nano Letters}\ }\textbf {\bibinfo {volume} {18}},\ \bibinfo {pages}
  {2033--2039} (\bibinfo {year} {2018})}\BibitemShut {NoStop}%
\bibitem [{\citenamefont {Li}\ \emph {et~al.}(2013)\citenamefont {Li},
  \citenamefont {Shao}, \citenamefont {Li}, \citenamefont {McCall},
  \citenamefont {Wang},\ and\ \citenamefont {Zhang}}]{Li2013}%
  \BibitemOpen
  \bibfield  {author} {\bibinfo {author} {\bibfnamefont {Z.}~\bibnamefont
  {Li}}, \bibinfo {author} {\bibfnamefont {S.}~\bibnamefont {Shao}}, \bibinfo
  {author} {\bibfnamefont {N.}~\bibnamefont {Li}}, \bibinfo {author}
  {\bibfnamefont {K.}~\bibnamefont {McCall}}, \bibinfo {author} {\bibfnamefont
  {J.}~\bibnamefont {Wang}}, \ and\ \bibinfo {author} {\bibfnamefont {S.~X.}\
  \bibnamefont {Zhang}},\ }\bibfield  {title} {\enquote {\bibinfo {title}
  {Single crystalline nanostructures of topological crystalline insulator snte
  with distinct facets and morphologies},}\ }\href {\doibase 10.1021/nl4030193}
  {\bibfield  {journal} {\bibinfo  {journal} {Nano Letters}\ }\textbf {\bibinfo
  {volume} {13}},\ \bibinfo {pages} {5443--5448} (\bibinfo {year} {2013})},\
  \Eprint {http://arxiv.org/abs/https://doi.org/10.1021/nl4030193}
  {https://doi.org/10.1021/nl4030193} \BibitemShut {NoStop}%
\bibitem [{\citenamefont {Safdar}\ \emph {et~al.}(2013)\citenamefont {Safdar},
  \citenamefont {Wang}, \citenamefont {Mirza}, \citenamefont {Wang},
  \citenamefont {Xu},\ and\ \citenamefont {He}}]{Safdar2013}%
  \BibitemOpen
  \bibfield  {author} {\bibinfo {author} {\bibfnamefont {Muhammad}\
  \bibnamefont {Safdar}}, \bibinfo {author} {\bibfnamefont {Qisheng}\
  \bibnamefont {Wang}}, \bibinfo {author} {\bibfnamefont {Misbah}\ \bibnamefont
  {Mirza}}, \bibinfo {author} {\bibfnamefont {Zhenxing}\ \bibnamefont {Wang}},
  \bibinfo {author} {\bibfnamefont {Kai}\ \bibnamefont {Xu}}, \ and\ \bibinfo
  {author} {\bibfnamefont {Jun}\ \bibnamefont {He}},\ }\bibfield  {title}
  {\enquote {\bibinfo {title} {Topological surface transport properties of
  single-crystalline snte nanowire},}\ }\href {\doibase 10.1021/nl402841x}
  {\bibfield  {journal} {\bibinfo  {journal} {Nano Letters}\ }\textbf {\bibinfo
  {volume} {13}},\ \bibinfo {pages} {5344--5349} (\bibinfo {year} {2013})},\
  \Eprint {http://arxiv.org/abs/https://doi.org/10.1021/nl402841x}
  {https://doi.org/10.1021/nl402841x} \BibitemShut {NoStop}%
\bibitem [{\citenamefont {Safdar}\ \emph {et~al.}(2014)\citenamefont {Safdar},
  \citenamefont {Wang}, \citenamefont {Mirza}, \citenamefont {Wang},\ and\
  \citenamefont {He}}]{Safdar2014}%
  \BibitemOpen
  \bibfield  {author} {\bibinfo {author} {\bibfnamefont {Muhammad}\
  \bibnamefont {Safdar}}, \bibinfo {author} {\bibfnamefont {Qisheng}\
  \bibnamefont {Wang}}, \bibinfo {author} {\bibfnamefont {Misbah}\ \bibnamefont
  {Mirza}}, \bibinfo {author} {\bibfnamefont {Zhenxing}\ \bibnamefont {Wang}},
  \ and\ \bibinfo {author} {\bibfnamefont {Jun}\ \bibnamefont {He}},\
  }\bibfield  {title} {\enquote {\bibinfo {title} {Crystal shape engineering of
  topological crystalline insulator snte microcrystals and nanowires with huge
  thermal activation energy gap},}\ }\href {\doibase 10.1021/cg5002122}
  {\bibfield  {journal} {\bibinfo  {journal} {Crystal Growth \& Design}\
  }\textbf {\bibinfo {volume} {14}},\ \bibinfo {pages} {2502--2509} (\bibinfo
  {year} {2014})}\BibitemShut {NoStop}%
\bibitem [{\citenamefont {Shen}\ \emph {et~al.}(2014)\citenamefont {Shen},
  \citenamefont {Jung}, \citenamefont {Disa}, \citenamefont {Walker},
  \citenamefont {Ahn},\ and\ \citenamefont {Cha}}]{Shen2014}%
  \BibitemOpen
  \bibfield  {author} {\bibinfo {author} {\bibfnamefont {Jie}\ \bibnamefont
  {Shen}}, \bibinfo {author} {\bibfnamefont {Yeonwoong}\ \bibnamefont {Jung}},
  \bibinfo {author} {\bibfnamefont {Ankit~S.}\ \bibnamefont {Disa}}, \bibinfo
  {author} {\bibfnamefont {Fred~J.}\ \bibnamefont {Walker}}, \bibinfo {author}
  {\bibfnamefont {Charles~H.}\ \bibnamefont {Ahn}}, \ and\ \bibinfo {author}
  {\bibfnamefont {Judy~J.}\ \bibnamefont {Cha}},\ }\bibfield  {title} {\enquote
  {\bibinfo {title} {Synthesis of snte nanoplates with \{100\} and \{111\}
  surfaces},}\ }\href {\doibase 10.1021/nl501953s} {\bibfield  {journal}
  {\bibinfo  {journal} {Nano Letters}\ }\textbf {\bibinfo {volume} {14}},\
  \bibinfo {pages} {4183--4188} (\bibinfo {year} {2014})}\BibitemShut {NoStop}%
\bibitem [{\citenamefont {Zou}\ \emph {et~al.}(2015)\citenamefont {Zou},
  \citenamefont {Chen}, \citenamefont {Lin}, \citenamefont {Zhou},
  \citenamefont {Lu}, \citenamefont {Drennan},\ and\ \citenamefont
  {Zou}}]{Zou2015}%
  \BibitemOpen
  \bibfield  {author} {\bibinfo {author} {\bibfnamefont {Yichao}\ \bibnamefont
  {Zou}}, \bibinfo {author} {\bibfnamefont {Zhigang}\ \bibnamefont {Chen}},
  \bibinfo {author} {\bibfnamefont {Jing}\ \bibnamefont {Lin}}, \bibinfo
  {author} {\bibfnamefont {Xiaohao}\ \bibnamefont {Zhou}}, \bibinfo {author}
  {\bibfnamefont {Wei}\ \bibnamefont {Lu}}, \bibinfo {author} {\bibfnamefont
  {John}\ \bibnamefont {Drennan}}, \ and\ \bibinfo {author} {\bibfnamefont
  {Jin}\ \bibnamefont {Zou}},\ }\bibfield  {title} {\enquote {\bibinfo {title}
  {Morphological control of snte nanostructures by tuning catalyst
  composition},}\ }\href {\doibase 10.1007/s12274-015-0806-y} {\bibfield
  {journal} {\bibinfo  {journal} {Nano Research}\ }\textbf {\bibinfo {volume}
  {8}},\ \bibinfo {pages} {3011--3019} (\bibinfo {year} {2015})}\BibitemShut
  {NoStop}%
\bibitem [{\citenamefont {Mata}\ \emph {et~al.}(2013)\citenamefont {Mata},
  \citenamefont {Zhou}, \citenamefont {Furtmayr}, \citenamefont {Teubert},
  \citenamefont {Gradecak}, \citenamefont {Eickhoff}, \citenamefont
  {Fontcuberta~i Morral},\ and\ \citenamefont {Arbiol}}]{Mata2013}%
  \BibitemOpen
  \bibfield  {author} {\bibinfo {author} {\bibfnamefont {Maria de~la}\
  \bibnamefont {Mata}}, \bibinfo {author} {\bibfnamefont {Xiang}\ \bibnamefont
  {Zhou}}, \bibinfo {author} {\bibfnamefont {Florian}\ \bibnamefont
  {Furtmayr}}, \bibinfo {author} {\bibfnamefont {Jorg}\ \bibnamefont
  {Teubert}}, \bibinfo {author} {\bibfnamefont {Silvija}\ \bibnamefont
  {Gradecak}}, \bibinfo {author} {\bibfnamefont {Martin}\ \bibnamefont
  {Eickhoff}}, \bibinfo {author} {\bibfnamefont {Anna}\ \bibnamefont
  {Fontcuberta~i Morral}}, \ and\ \bibinfo {author} {\bibfnamefont {Jordi}\
  \bibnamefont {Arbiol}},\ }\bibfield  {title} {\enquote {\bibinfo {title} {A
  review of mbe grown 0d{,} 1d and 2d quantum structures in a nanowire},}\
  }\href {\doibase 10.1039/C3TC30556B} {\bibfield  {journal} {\bibinfo
  {journal} {J. Mater. Chem. C}\ }\textbf {\bibinfo {volume} {1}},\ \bibinfo
  {pages} {4300--4312} (\bibinfo {year} {2013})}\BibitemShut {NoStop}%
\bibitem [{\citenamefont {Strupinski}\ \emph {et~al.}(2011)\citenamefont
  {Strupinski}, \citenamefont {Grodecki}, \citenamefont {Wysmolek},
  \citenamefont {Stepniewski}, \citenamefont {Szkopek}, \citenamefont
  {Gaskell}, \citenamefont {Gr{\"u}neis}, \citenamefont {Haberer},
  \citenamefont {Bozek}, \citenamefont {Krupka},\ and\ \citenamefont
  {Baranowski}}]{Strupinski2011}%
  \BibitemOpen
  \bibfield  {author} {\bibinfo {author} {\bibfnamefont {W.}~\bibnamefont
  {Strupinski}}, \bibinfo {author} {\bibfnamefont {K.}~\bibnamefont
  {Grodecki}}, \bibinfo {author} {\bibfnamefont {A.}~\bibnamefont {Wysmolek}},
  \bibinfo {author} {\bibfnamefont {R.}~\bibnamefont {Stepniewski}}, \bibinfo
  {author} {\bibfnamefont {T.}~\bibnamefont {Szkopek}}, \bibinfo {author}
  {\bibfnamefont {P.~E.}\ \bibnamefont {Gaskell}}, \bibinfo {author}
  {\bibfnamefont {A.}~\bibnamefont {Gr{\"u}neis}}, \bibinfo {author}
  {\bibfnamefont {D.}~\bibnamefont {Haberer}}, \bibinfo {author} {\bibfnamefont
  {R.}~\bibnamefont {Bozek}}, \bibinfo {author} {\bibfnamefont
  {J.}~\bibnamefont {Krupka}}, \ and\ \bibinfo {author} {\bibfnamefont {J.~M.}\
  \bibnamefont {Baranowski}},\ }\bibfield  {title} {\enquote {\bibinfo {title}
  {Graphene epitaxy by chemical vapor deposition on sic},}\ }\href {\doibase
  10.1021/nl200390e} {\bibfield  {journal} {\bibinfo  {journal} {Nano Letters}\
  }\textbf {\bibinfo {volume} {11}},\ \bibinfo {pages} {1786--1791} (\bibinfo
  {year} {2011})}\BibitemShut {NoStop}%
\bibitem [{\citenamefont {Dziawa}\ \emph {et~al.}(2010)\citenamefont {Dziawa},
  \citenamefont {Sadowski}, \citenamefont {Dluzewski}, \citenamefont
  {Lusakowska}, \citenamefont {Domukhovski}, \citenamefont {Taliashvili},
  \citenamefont {Wojciechowski}, \citenamefont {Baczewski}, \citenamefont
  {Bukala}, \citenamefont {Galicka}, \citenamefont {Buczko}, \citenamefont
  {Kacman},\ and\ \citenamefont {Story}}]{Dziawa2010}%
  \BibitemOpen
  \bibfield  {author} {\bibinfo {author} {\bibfnamefont {P.}~\bibnamefont
  {Dziawa}}, \bibinfo {author} {\bibfnamefont {J.}~\bibnamefont {Sadowski}},
  \bibinfo {author} {\bibfnamefont {P.}~\bibnamefont {Dluzewski}}, \bibinfo
  {author} {\bibfnamefont {E.}~\bibnamefont {Lusakowska}}, \bibinfo {author}
  {\bibfnamefont {V.}~\bibnamefont {Domukhovski}}, \bibinfo {author}
  {\bibfnamefont {B.}~\bibnamefont {Taliashvili}}, \bibinfo {author}
  {\bibfnamefont {T.}~\bibnamefont {Wojciechowski}}, \bibinfo {author}
  {\bibfnamefont {L.~T.}\ \bibnamefont {Baczewski}}, \bibinfo {author}
  {\bibfnamefont {M.}~\bibnamefont {Bukala}}, \bibinfo {author} {\bibfnamefont
  {M.}~\bibnamefont {Galicka}}, \bibinfo {author} {\bibfnamefont
  {R.}~\bibnamefont {Buczko}}, \bibinfo {author} {\bibfnamefont
  {P.}~\bibnamefont {Kacman}}, \ and\ \bibinfo {author} {\bibfnamefont
  {T.}~\bibnamefont {Story}},\ }\bibfield  {title} {\enquote {\bibinfo {title}
  {Defect free pbte nanowires grown by molecular beam epitaxy on gaas(111)b
  substrates},}\ }\href {\doibase 10.1021/cg900575r} {\bibfield  {journal}
  {\bibinfo  {journal} {Crystal Growth \& Design}\ }\textbf {\bibinfo {volume}
  {10}},\ \bibinfo {pages} {109--113} (\bibinfo {year} {2010})}\BibitemShut
  {NoStop}%
\bibitem [{\citenamefont {Jia}\ \emph {et~al.}(2010)\citenamefont {Jia},
  \citenamefont {Houben}, \citenamefont {Thust},\ and\ \citenamefont
  {Barthel}}]{Jia2010}%
  \BibitemOpen
  \bibfield  {author} {\bibinfo {author} {\bibfnamefont {C.~L.}\ \bibnamefont
  {Jia}}, \bibinfo {author} {\bibfnamefont {L.}~\bibnamefont {Houben}},
  \bibinfo {author} {\bibfnamefont {A.}~\bibnamefont {Thust}}, \ and\ \bibinfo
  {author} {\bibfnamefont {J.}~\bibnamefont {Barthel}},\ }\bibfield  {title}
  {\enquote {\bibinfo {title} {On the benefit of the
  negative-spherical-aberration imaging technique for quantitative hrtem},}\
  }\href {http://www.sciencedirect.com/science/article/pii/S030439910900223X}
  {\bibfield  {journal} {\bibinfo  {journal} {Ultramicroscopy}\ }\textbf
  {\bibinfo {volume} {110}},\ \bibinfo {pages} {500--505} (\bibinfo {year}
  {2010})}\BibitemShut {NoStop}%
\bibitem [{\citenamefont {Norimatsu}\ and\ \citenamefont
  {Kusunoki}(2009)}]{Norimatsu2009}%
  \BibitemOpen
  \bibfield  {author} {\bibinfo {author} {\bibfnamefont {Wataru}\ \bibnamefont
  {Norimatsu}}\ and\ \bibinfo {author} {\bibfnamefont {Michiko}\ \bibnamefont
  {Kusunoki}},\ }\href@noop {} {\emph {\bibinfo {title} {Transitional
  structures of the interface between graphene and 6H-SiC (0 0 0 1)}}},\
  \bibinfo {series} {Chemical Physics Letters - CHEM PHYS LETT}, Vol.\ \bibinfo
  {volume} {468}\ (\bibinfo {year} {2009})\ pp.\ \bibinfo {pages}
  {52--56}\BibitemShut {NoStop}%
\bibitem [{\citenamefont {Chang}\ \emph {et~al.}(2016)\citenamefont {Chang},
  \citenamefont {Liu}, \citenamefont {Lin}, \citenamefont {Wang}, \citenamefont
  {Zhao}, \citenamefont {Zhang}, \citenamefont {Jin}, \citenamefont {Zhong},
  \citenamefont {Hu}, \citenamefont {Duan}, \citenamefont {Zhang},
  \citenamefont {Fu}, \citenamefont {Xue}, \citenamefont {Chen},\ and\
  \citenamefont {Ji}}]{Chang2016}%
  \BibitemOpen
  \bibfield  {author} {\bibinfo {author} {\bibfnamefont {Kai}\ \bibnamefont
  {Chang}}, \bibinfo {author} {\bibfnamefont {Junwei}\ \bibnamefont {Liu}},
  \bibinfo {author} {\bibfnamefont {Haicheng}\ \bibnamefont {Lin}}, \bibinfo
  {author} {\bibfnamefont {Na}~\bibnamefont {Wang}}, \bibinfo {author}
  {\bibfnamefont {Kun}\ \bibnamefont {Zhao}}, \bibinfo {author} {\bibfnamefont
  {Anmin}\ \bibnamefont {Zhang}}, \bibinfo {author} {\bibfnamefont {Feng}\
  \bibnamefont {Jin}}, \bibinfo {author} {\bibfnamefont {Yong}\ \bibnamefont
  {Zhong}}, \bibinfo {author} {\bibfnamefont {Xiaopeng}\ \bibnamefont {Hu}},
  \bibinfo {author} {\bibfnamefont {Wenhui}\ \bibnamefont {Duan}}, \bibinfo
  {author} {\bibfnamefont {Qingming}\ \bibnamefont {Zhang}}, \bibinfo {author}
  {\bibfnamefont {Liang}\ \bibnamefont {Fu}}, \bibinfo {author} {\bibfnamefont
  {Qi-Kun}\ \bibnamefont {Xue}}, \bibinfo {author} {\bibfnamefont
  {Xi}~\bibnamefont {Chen}}, \ and\ \bibinfo {author} {\bibfnamefont
  {Shuai-Hua}\ \bibnamefont {Ji}},\ }\bibfield  {title} {\enquote {\bibinfo
  {title} {Discovery of robust in-plane ferroelectricity in atomic-thick
  snte},}\ }\href {\doibase 10.1126/science.aad8609} {\bibfield  {journal}
  {\bibinfo  {journal} {Science}\ }\textbf {\bibinfo {volume} {353}},\ \bibinfo
  {pages} {274} (\bibinfo {year} {2016})}\BibitemShut {NoStop}%
\bibitem [{\citenamefont {Munshi}\ \emph {et~al.}(2012)\citenamefont {Munshi},
  \citenamefont {Dheeraj}, \citenamefont {Fauske}, \citenamefont {Kim},
  \citenamefont {van Helvoort}, \citenamefont {Fimland},\ and\ \citenamefont
  {Weman}}]{Munshi2012}%
  \BibitemOpen
  \bibfield  {author} {\bibinfo {author} {\bibfnamefont {A.~Mazid}\
  \bibnamefont {Munshi}}, \bibinfo {author} {\bibfnamefont {Dasa~L.}\
  \bibnamefont {Dheeraj}}, \bibinfo {author} {\bibfnamefont {Vidar~T.}\
  \bibnamefont {Fauske}}, \bibinfo {author} {\bibfnamefont {Dong-Chul}\
  \bibnamefont {Kim}}, \bibinfo {author} {\bibfnamefont {Antonius T.~J.}\
  \bibnamefont {van Helvoort}}, \bibinfo {author} {\bibfnamefont
  {Bj{\o}rn-Ove}\ \bibnamefont {Fimland}}, \ and\ \bibinfo {author}
  {\bibfnamefont {Helge}\ \bibnamefont {Weman}},\ }\bibfield  {title} {\enquote
  {\bibinfo {title} {Vertically aligned gaas nanowires on graphite and
  few-layer graphene: Generic model and epitaxial growth},}\ }\href {\doibase
  10.1021/nl3018115} {\bibfield  {journal} {\bibinfo  {journal} {Nano Letters}\
  }\textbf {\bibinfo {volume} {12}},\ \bibinfo {pages} {4570--4576} (\bibinfo
  {year} {2012})}\BibitemShut {NoStop}%
\bibitem [{\citenamefont {Arora}\ and\ \citenamefont
  {Jagirdar}(2014)}]{Arora2014}%
  \BibitemOpen
  \bibfield  {author} {\bibinfo {author} {\bibfnamefont {Neha}\ \bibnamefont
  {Arora}}\ and\ \bibinfo {author} {\bibfnamefont {Balaji~R.}\ \bibnamefont
  {Jagirdar}},\ }\bibfield  {title} {\enquote {\bibinfo {title} {From (au5sn +
  ausn) physical mixture to phase pure ausn and au5sn intermetallic
  nanocrystals with tailored morphology: digestive ripening assisted
  approach},}\ }\href {\doibase 10.1039/C4CP00249K} {\bibfield  {journal}
  {\bibinfo  {journal} {Physical Chemistry Chemical Physics}\ }\textbf
  {\bibinfo {volume} {16}},\ \bibinfo {pages} {11381--11389} (\bibinfo {year}
  {2014})}\BibitemShut {NoStop}%
\bibitem [{\citenamefont {Berchenko}\ \emph {et~al.}(2018)\citenamefont
  {Berchenko}, \citenamefont {Vitchev}, \citenamefont {Trzyna}, \citenamefont
  {Wojnarowska-Nowak}, \citenamefont {Szczerbakow}, \citenamefont {Badyla},
  \citenamefont {Cebulski},\ and\ \citenamefont {Story}}]{Berchenko2018}%
  \BibitemOpen
  \bibfield  {author} {\bibinfo {author} {\bibfnamefont {N.}~\bibnamefont
  {Berchenko}}, \bibinfo {author} {\bibfnamefont {R.}~\bibnamefont {Vitchev}},
  \bibinfo {author} {\bibfnamefont {M.}~\bibnamefont {Trzyna}}, \bibinfo
  {author} {\bibfnamefont {R.}~\bibnamefont {Wojnarowska-Nowak}}, \bibinfo
  {author} {\bibfnamefont {A.}~\bibnamefont {Szczerbakow}}, \bibinfo {author}
  {\bibfnamefont {A.}~\bibnamefont {Badyla}}, \bibinfo {author} {\bibfnamefont
  {J.}~\bibnamefont {Cebulski}}, \ and\ \bibinfo {author} {\bibfnamefont
  {T.}~\bibnamefont {Story}},\ }\bibfield  {title} {\enquote {\bibinfo {title}
  {Surface oxidation of snte topological crystalline insulator},}\ }\href
  {http://www.sciencedirect.com/science/article/pii/S0169433218312261}
  {\bibfield  {journal} {\bibinfo  {journal} {Applied Surface Science}\
  }\textbf {\bibinfo {volume} {452}},\ \bibinfo {pages} {134--140} (\bibinfo
  {year} {2018})}\BibitemShut {NoStop}%
\bibitem [{\citenamefont {Brebrick}\ and\ \citenamefont
  {Strauss}(1964)}]{Brebrick1964}%
  \BibitemOpen
  \bibfield  {author} {\bibinfo {author} {\bibfnamefont {R.~F.}\ \bibnamefont
  {Brebrick}}\ and\ \bibinfo {author} {\bibfnamefont {A.~J.}\ \bibnamefont
  {Strauss}},\ }\bibfield  {title} {\enquote {\bibinfo {title} {Partial
  pressures in equilibrium with group iv tellurides. ii. tin telluride},}\
  }\href {\doibase 10.1063/1.1725623} {\bibfield  {journal} {\bibinfo
  {journal} {The Journal of Chemical Physics}\ }\textbf {\bibinfo {volume}
  {41}},\ \bibinfo {pages} {197--205} (\bibinfo {year} {1964})}\BibitemShut
  {NoStop}%
\bibitem [{\citenamefont {Okamoto}\ and\ \citenamefont
  {Massalski}(1984)}]{Okamoto1984}%
  \BibitemOpen
  \bibfield  {author} {\bibinfo {author} {\bibfnamefont {H.}~\bibnamefont
  {Okamoto}}\ and\ \bibinfo {author} {\bibfnamefont {T.~B.}\ \bibnamefont
  {Massalski}},\ }\bibfield  {title} {\enquote {\bibinfo {title} {The au-sn
  (gold-tin) system},}\ }\href {\doibase 10.1007/BF02872904} {\bibfield
  {journal} {\bibinfo  {journal} {Bulletin of Alloy Phase Diagrams}\ }\textbf
  {\bibinfo {volume} {5}},\ \bibinfo {pages} {492} (\bibinfo {year}
  {1984})}\BibitemShut {NoStop}%
\bibitem [{\citenamefont {Colliex}\ \emph {et~al.}(2016)\citenamefont
  {Colliex}, \citenamefont {Kociak},\ and\ \citenamefont
  {St{\'e}phan}}]{Colliex2016}%
  \BibitemOpen
  \bibfield  {author} {\bibinfo {author} {\bibfnamefont {Christian}\
  \bibnamefont {Colliex}}, \bibinfo {author} {\bibfnamefont {Mathieu}\
  \bibnamefont {Kociak}}, \ and\ \bibinfo {author} {\bibfnamefont {Odile}\
  \bibnamefont {St{\'e}phan}},\ }\bibfield  {title} {\enquote {\bibinfo {title}
  {Electron energy loss spectroscopy imaging of surface plasmons at the
  nanometer scale},}\ }\href
  {http://www.sciencedirect.com/science/article/pii/S0304399115300760}
  {\bibfield  {journal} {\bibinfo  {journal} {Ultramicroscopy}\ }\textbf
  {\bibinfo {volume} {162}},\ \bibinfo {pages} {A1--A24} (\bibinfo {year}
  {2016})}\BibitemShut {NoStop}%
\bibitem [{\citenamefont {Bosman}\ \emph {et~al.}(2007)\citenamefont {Bosman},
  \citenamefont {Keast}, \citenamefont {Watanabe}, \citenamefont {Maaroof},\
  and\ \citenamefont {Cortie}}]{Bosman2007}%
  \BibitemOpen
  \bibfield  {author} {\bibinfo {author} {\bibfnamefont {Michel}\ \bibnamefont
  {Bosman}}, \bibinfo {author} {\bibfnamefont {Vicki~J}\ \bibnamefont {Keast}},
  \bibinfo {author} {\bibfnamefont {Masashi}\ \bibnamefont {Watanabe}},
  \bibinfo {author} {\bibfnamefont {Abbas~I}\ \bibnamefont {Maaroof}}, \ and\
  \bibinfo {author} {\bibfnamefont {Michael~B}\ \bibnamefont {Cortie}},\
  }\bibfield  {title} {\enquote {\bibinfo {title} {Mapping surface plasmons at
  the nanometre scale with an electron beam},}\ }\href
  {http://stacks.iop.org/0957-4484/18/i=16/a=165505} {\bibfield  {journal}
  {\bibinfo  {journal} {Nanotechnology}\ }\textbf {\bibinfo {volume} {18}},\
  \bibinfo {pages} {165505} (\bibinfo {year} {2007})}\BibitemShut {NoStop}%
\bibitem [{\citenamefont {Cook}(1971)}]{Cook1971}%
  \BibitemOpen
  \bibfield  {author} {\bibinfo {author} {\bibfnamefont {R.~F.}\ \bibnamefont
  {Cook}},\ }\bibfield  {title} {\enquote {\bibinfo {title} {Electron energy
  loss spectra of gete-snte alloys},}\ }\href@noop {} {\bibfield  {journal}
  {\bibinfo  {journal} {Philosophical Magazine}\ }\textbf {\bibinfo {volume}
  {24}},\ \bibinfo {pages} {1347--1353} (\bibinfo {year} {1971})}\BibitemShut
  {NoStop}%
\end{thebibliography}%

\end{document}